\documentstyle[prl,aps,epsfig,preprint]{revtex}
\tighten
\begin{document}

\title{Macroscopic entanglement in Josephson nanocircuits}
\author{Francesco Plastina $^{(1,3)}$, Rosario Fazio$^{(2,3)}$,
    and G. Massimo Palma $^{(1,3)}$}
\address{$^{(1)}$Dipartimento di Scienze Fisiche e
Astronomiche (DSFA)
    Universit\`a di Palermo, via Archirafi 36, I-90123 Palermo\\
    $^{(2)}$ Dipartimento di Metodologie Fisiche
    e Chimiche (DMFCI),Universit\`a di Catania,
    viale A.Doria 6, I-95125 Catania, Italy\\
    $^{(3)}$Istituto Nazionale di Fisica della Materia, Unit\`a di
    Catania e Palermo }

\date{\today}
\maketitle

\begin{abstract}
In this Letter we propose a scheme to generate and detect entanglement between
charge states in superconducting nanocircuits.  We discuss different
procedures to discriminate such entanglement from
classical correlations. The case of maximally entangled states of
two and three coupled Josephson junctions is discussed as example.
\end{abstract}



The phenomenon of entanglement, probably one of the most striking
feature of quantum mechanics~\cite{entanglement}, appearing as a
consequence of the superposition principle in the presence of
composite systems, is the main ingredient in all
known examples of quantum speed-up in quantum computation and
communication~\cite{jozsa,nielsen}. This has prompted intense
experimental efforts towards its generation and detection,
most notably with photons~\cite{zeilinger}, cavity QED
systems~\cite{haroche}, ion traps \cite{wineland}, and coupled
quantum dots~\cite{bayer}. Although  in condensed matter it is
common to encounter correlated many-body states, it is difficult to
isolate the different subsystems while maintaining their entanglement.
This problem is recently attracting a lot of attention and 
several solid state devices have been suggested. 
They are based on the phenomenon of Andreev reflection in 
hybrid normal-superconducting systems~\cite{loss,lesovik} or on
the coupling of mesoscopic Josephson junctions with superconducting
resonators~\cite{buisson,marquardt}.

In this Letter we propose an explicit experimental scheme to {\em generate} 
and {\em detect} entanglement in superconducting nanocircuits. 
Astonishing progresses have recently been made in the
control of the coherent evolution of such systems~\cite{bouchiat,nakamura}, 
which have been proposed as promising
candidate to realize a quantum computer~\cite{schoen,averin,fluxqubit,falci}.
In our proposal we specifically address two important issues. 
i)  We describe how to measure the entangled states once the subsystems have 
been separated and what is the effect of a possible residual coupling on the 
outcome of the measurements. 
ii) We show how to distinguish between
entangled states and statistical mixtures or product states. 
Our setup is based on a modification of the device used by Nakamura 
{\em et al.}~\cite{nakamura,choi} and, we believe, it is amenable of
experimental verification with present days technology.

\begin{figure}
\begin{center}
\epsfig{figure=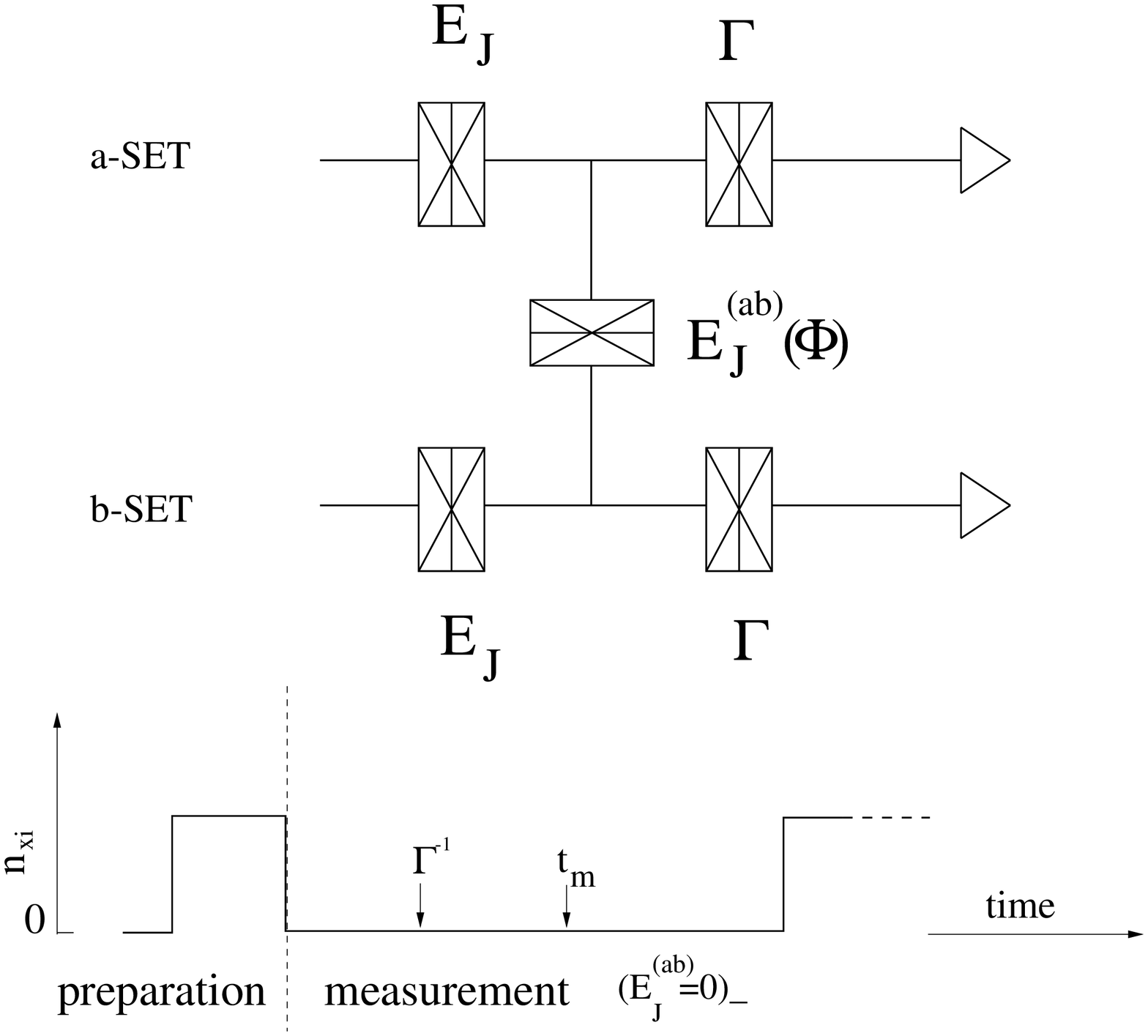,width=1\textwidth}
\end{center}
\caption{(Top) Schematic representation of the proposed setup. 
The left junctions and the SQUID allow for 
the coherent manipulations to prepare the state. The right junctions provide the 
measurement setup.
(Bottom) Procedure to prepare and detect the entanglement. 
The procedure in repeated as in the
experiment in Ref \protect\cite{nakamura}}
\label{uno}
\end{figure}

As sketched in Fig.\ref{uno}, we consider two superconducting Single 
Electron Tunneling (SET) transistors (labelled by $a,b$) coupled by a small, 
Josephson junction.
By choosing appropriately the working point of the device~\cite{nakamura} 
(see below), coherent Cooper-pair tunneling takes place only across the left 
and the coupling junctions, while quasi-particle tunneling 
is important across the right junctions. 
With these observations the Hamiltonian can be written as a sum of three contributions
$H=H_0+H_{qp}+H_T$. Its coherent part, $H_0$, is given by
\begin{eqnarray}
    H_0  &=&
    \sum_{i=a,b } \left[  E_{ch}
    (n_{i}-n_{x,i})^2
    - E_{ J}  \cos 2 \varphi_{i} \right]  \nonumber \\
    &+& E^{(ab)}_{c} (n_a-n_{x,a}) (n_b-n_{x,b}) +
    E_{J}^{(ab)}(\Phi ) \cos 2(\varphi _a - \varphi _b ) \label{ham}
\end{eqnarray}
Here, $E_{ch}$ is the charging energy, $n_{x,i}$ are offset charges 
induced by external voltages and $E_{J}$ is associated to the Josephson 
tunneling. The phases $\varphi_i$ and the number of charges on the islands  
$n_i$  are conjugate variables 
$[\varphi_{i},n_{j}]= \,i \; \delta_{ij}$. 
The state of each individual SET can be manipulated by varying $n_x$ 
with a suitable choice of voltage pulses, for example by putting the left 
junctions at resonance for Cooper-pair tunneling. However, in order to produce 
entanglement, we must be able to perform transformations other than local ones.
This requires a controllable coupling between the SETs. In our proposal such 
coupling, $E_{J}^{(ab)}(\Phi )$,  is provided by a SQUID, pierced by an external 
flux $\Phi$. The SQUID capacitance, which we assume to be much smaller than 
all the other capacitances in the setup, gives rise to an electrostatic coupling, 
$E^{(ab)}_{c}$. The term
$H_{qp}= \sum_{\alpha}\sum_{k\sigma} \varepsilon_{k\alpha}
\gamma_{k\sigma\alpha}^\dag \gamma_{k\sigma \alpha}^{}$
($\alpha=R_a,R_b,I_a,I_b$ labels the right electrodes and both islands respectively) 
describes the quasi-particles. Here, $\gamma_{k \sigma \alpha}^\dag$ 
($\gamma_{k \sigma \alpha}$) creates (destroyes) a quasiparticle with momentum $k$ 
and energy
$\varepsilon_{k\alpha} = \sqrt{\xi_{k\alpha}^2+\Delta^2}$ 
($\xi_k$ is the single-particle dispersion, $\Delta$ is the superconducting
gap and $\sigma$ the spin label).
Finally the tunneling Hamiltonian $H_T$ is 
$H_T = \sum_{i=a,b} \left[ e^{-i\varphi_i}
	\sum_{kq\sigma}T_{kq}\gamma_{k\sigma R_i}^\dag \gamma_{q\sigma I_i}
	+ h.c.\right]$  where
$T_{kq}$ is the amplitude for quasi-particle tunneling. By fixing the transport 
voltage $|eV|\simeq 2\Delta + E_{ch}$ and due to the Coulomb blockade 
$E_{ch} \gg E_J$, it suffices to consider only three charge states, $n=0,1,2$ and 
two quasi-particle tunneling rates, $\Gamma_2^{a/b}$ and 
$\Gamma_1^{a/b}$. The latter are related to transitions across the right junctions 
with the charge on the island changing as $n=2\to n=1$ and $n=1\to n=0$ 
respectively (in the regime we consider, $\Gamma_2^{(a/b)} \sim \Gamma_1^{(a/b)}\sim
\Gamma$ )~\cite{aleshkin}. The other tunneling rates are exponentially small. 
We furthermore assume $\Gamma \ll E_J$.

The dynamics of the system can be described by a master equation 
for the density matrix representing the charge state of the two 
islands~\cite{aleshkin}
\begin{equation}
    \partial_t \rho = i \Bigl [\rho, H_0 \Bigr ] 
	+
     \frac{1}{2} \sum_{i=a,b}
    \sum_{n=1,2} \Gamma_n^{(i)} \biggl \{ 2 L_n^{(i)} \rho
    \Bigl (L_n^{(i)} \Bigr )^{\dag}
    -\Bigl (L_n^{(i)} \Bigr )^{\dag} L_n^{(i)} \rho -
    \rho \Bigl (L_n^{(i)} \Bigr )^{\dag}L_n^{(i)} \biggr \}
\label{master}
\end{equation}
\noindent
where $L_n^{(i)}=|n-1\rangle\langle n|_i$ is the Lindblad operator corresponding 
to the quantum jump $n\rightarrow n-1$ for the $i$-th island. The states  
$|0\rangle$ and $|2\rangle$ of the two SETs are involved in generation of 
entanglement, while quasi-particle transitions through the states  $|1\rangle$ 
allow to  perform the quantum measurement.  
When such devices are employed as qubits  states $|0\rangle$ and $|2\rangle$  
are the computational states.

Ideally the detection of entanglement goes through the following
steps. i) Prepare the entangled
state by means of manipulation of the gate voltages; ii) Switch
off the coupling between the two qubits; iii) Perform the
measurement.

{\em Preparation} - We illustrate one of the possible procedures to prepare 
the singlet state,
$ |\Psi^{(-)}\rangle = (|02\rangle -|20\rangle) / \sqrt 2$. 
The two junctions are initially kept off degeneracy with the initial state
given by $|02\rangle$ . Then, by
switching on the Josephson coupling $E_J^{(ab)}$, and slightly shifting the 
working point of the two qubits such
that $n_{x,a} - n_{x,b} \simeq E_J^{(ab)} /E_{ch}$, for a time 
$\delta t = \pi/(\sqrt{2}E^{(ab)}_J)$, the desired
singlet state is obtained. In a similar fashion it is possible to generate 
other maximally entangled states. After
the preparation, the coupling between the two SETs should be switched off.

Equally important to the ability to generate entanglement is the possibility 
to detect it. Below we discuss two possible detection schemes.

{\em Method 1} - The success in the preparation of $|\Psi^{(-)}\rangle $ 
against  decoherence can be tested, as proposed in Ref.\cite{loss1}, by
measuring the correlation between the integrated quasiparticle current signals 
of the two SETs, $\tilde{I}_{a/b}=\int_0^{t_m}I_{a/b}(t) \, dt$, with the 
integration time, $t_m$, chosen to be much longer than
$\Gamma^{-1}$. The correlator is given by ($e$ is the electron charge)
\begin{equation}
{\cal C} = e^{-2}\langle \tilde I_a \tilde I_b \rangle 
	 = e^{-2} \int_0^{t_{m}} dt_1 \- \int_0^{t_{m}} dt_2 \, \langle I_a(t_1)
	   I_b(t_2) \rangle
\end{equation}
where the two-time average $\langle I_a(t_1) I_b(t_2) \rangle$ can be obtained 
with the help of the quantum regression theorem applied to eq. (\ref{master}). 
Following the procedure of Nakamura {\em et al.}~\cite{nakamura}, the experiment 
can be performed by means of repeated preparations of the initial state and
measurements of the correlator. A value of  ${\cal C}$ different from zero 
indicates  
that current is flowing in both channels. This cannot happen if the system is 
prepared in the $|\Psi^{(-)}\rangle$ state: if island $a$ relaxes from 
$|2\rangle$ to $|0\rangle$, then $b$ does not and viceversa.

Complete anticorrelation between the two currents is not enough to come 
to the conclusion that the charge state of the two islands is entangled. 
A statistical mixture of the form 
$\rho_{mix} = \frac{1}{2} ( |02 \rangle \langle 02 | + |20 \rangle \langle 20 | )$ 
or any convex combination of the kind $\rho_p=p \rho_{mix} + (1-p)
|\Psi^{(-)}\rangle \langle \Psi^{(-)} |$ would also lead to the same result for 
${\cal C}$. In order to discriminate between quantum vs. classical correlations 
we can proceed in the spirit of experiments testing Bell's
inequality~\cite{entanglement}. The basic idea is that  if the system is in 
state $|\Psi^{(-)}\rangle $ the result ${\cal C}=0$ is obtained also if, before 
the detection stage, one carries out a (further) bi--local unitary
operation. For example, one can rotate the state of the two qubits by the same 
angle $\theta$ bringing both islands to degeneracy for a time interval 
$\Delta t = 2 \theta /E_J$. This leaves $|\Psi^{(-)}\rangle $ unaffected. On the
other hand when the system is in a state like $\rho_{mix}$ bi--local unitary 
transformations will in general produce a non--zero population in $|22\rangle$ 
therefore giving rise to a non--zero ${\cal C}$. This is the key
ingredient of our scheme to discriminate entangled states from classical mixture.

In the ideal case we can think to switch off the Josephson interaction energy  
during the measurement, thus "separating" the two subsystems. If this can be 
done and if, in addition, also the Josephson energies of the two
transistors can be set to zero, then the master equation can be solved exactly. 
Starting with the state $\rho_p$, the integrated current correlation is given by
\begin{equation}
	{\cal C}_{p} 
	=  p \sin^2(2 \theta) 
	\left [1-e^{-\Gamma t_m}- \frac{\Gamma t_m}{2}e^{-\Gamma t_m} \right ]
\label{corr}
\end{equation}
The two cases of the singlet state and of the complete mixture are obtained for 
$p=0$ and $1$, respectively.  Note that for $p=1$ an oscillatory behaviour is 
obtained, whose visibility is reduced for $p<1$. Furthermore only when $p=0$
the correlator ${\cal C}$ vanishes. 

\begin{figure}
\begin{center}
\epsfig{figure=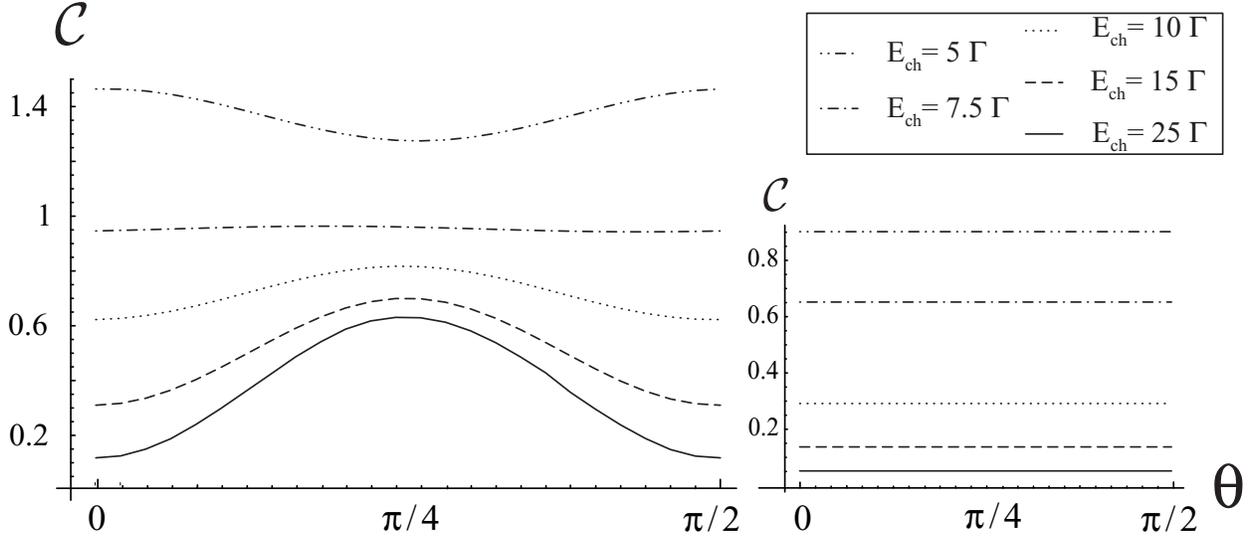,width=1\textwidth}
\end{center}
\caption{Time Average of the current correlation function over the
time interval $\Delta t=3 \Gamma^{-1}$  as a 
function of the rotating angle $\theta$ for
the case of the mixture (left) and the singlet (right).
The parameters in both figures are $E_J^{(ab)}=10 \Gamma$, $E_J=20 \Gamma$,
$E_c^{(ab)}=0.1 \Gamma$ and the indicated values for $E_{ch}$.} 
\label{C1}
\end{figure}

\begin{figure}
\begin{center}
\epsfig{figure=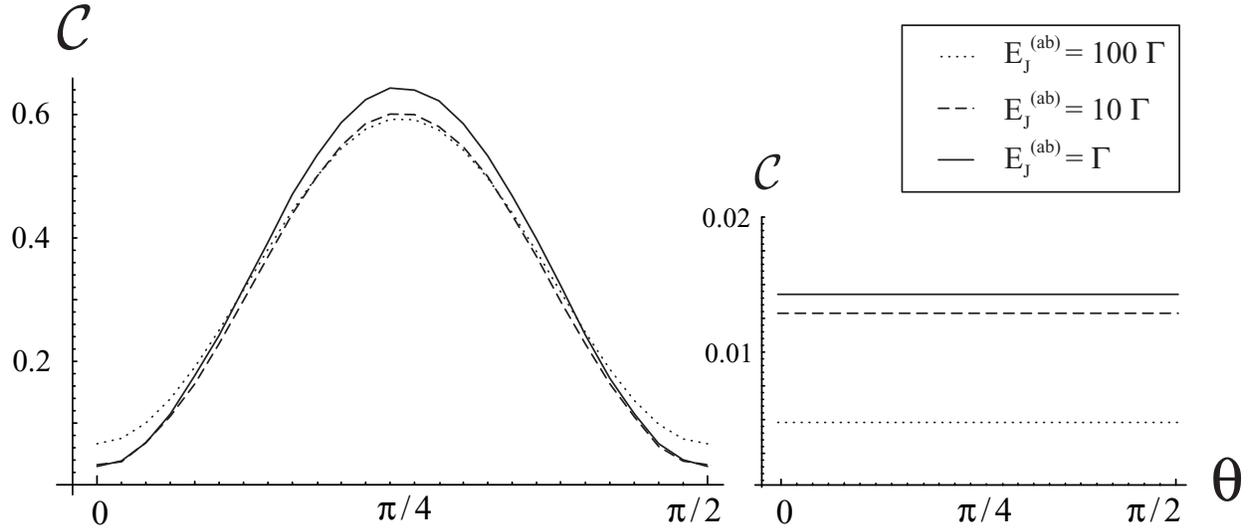,width=1\textwidth}
\end{center}
\caption{The same as in the previous figure, for different values of the 
Josephson coupling of the SQUID $E_{J}^{(ab)}$ 
($E_{ch}=25 \Gamma$, $E_J=10 \Gamma$, $E_c^{(ab)}=0.1 \Gamma$).
The correlator in both cases is almost insesitive to $E_{J}^{(ab)}$}
\label{C2}
\end{figure}

In order to ascertain the 
validity of this proposal it is important to check how the residual couplings 
affect  the  measurement. Experimentally 
a non-zero electrostatic coupling will be always present; however, as long as it is 
small compared to $E_{ch}$, it has no effect on the measurement. 
More important is the role of  $E_J$ and of the coupling $E_J^{(ab)}$; 
nevertheless, even if residual 
Josephson coupling energies are present during the measurement, the singlet and the ]
statistical mixture give rise to very different correlation signals.
In Figs.\ref{C1},\ref{C2} the correlation ${\cal C}$ is shown for 
$t_m=3 \Gamma^{-1}$, for both the $p=0$ and $p=1$ cases, and for different values of 
the charging energy of the SETs and $E_J^{(ab)}$, respectively.
During the measurement, the Cooper pair states $n=0,2$ should be kept 
off degeneracy. 
The amplitude of the oscillations in the mixed case depends on the charging energy 
$E_{ch}$ (for finite $E_J$) because the coherent oscillations 
in the two SETs may significantly change the state during the measurement.
Differences between  the pure and mixed state in Fig.\ref{C1} can be detected 
down to 
values of $E_{ch}/E_J \ge 1$ (see the dashed line in Fig.\ref{C1}). For lower values 
of  $E_{ch}/E_J$, it is impossible to discriminate between the $p=0$ and $p=1$ cases 
because of the extra Cooper pair tunneling.
It is theoretically desiderable and possible experimentally to fix the ratio 
$E_{ch}/E_J \gg 1$, therefore this is not a limitation for the feasibility of our 
proposal. The  signal does not show substantial variations with
increasing the Josephson coupling $E_J^{(ab)}$ either (see Fig.\ref{C2}).
The singlet state always gives rise to a correlation function which does not change 
with $\theta$. Additional errors, deriving from not perfect gating,
are possible; but they are not specific for the measurement of entanglement
and it was aready verified~\cite{nakamura} that they do not affect the preparation 
of the state.  One can conclude that our proposal 
is robust against various non-ideal situations that can be encountered in an
experiment.  

In order to perform current correlation measurements one needs detectors
sensitive to single electrons, which is technologically very demanding.
One possibility could be to amplify the signals coming from the two qubits before
correlating them. A similar procedure has been recently employed to measure
the correlation between the two output ports of a fermionic beam
splitter~\cite{yamamoto}.

{\em Method 2} - The protocol described so far is particularly suitable to 
verify that a given maximally entangled state has been created and to detect if 
decoherence mechanisms have spoiled  its coherence. A different approach,
which does not require the technologically demanding  current correlation 
measurement (nor an a priori knowledge of which state has to be detected) 
can be designed as well. The basic requirement of this approach is the ability to
implement a two--qubit gate, \cite{nota}.

To this end we consider the following sequence of voltage pulses. 
i) Bring the $b$-island to degeneracy with a
voltage pulse of duration $t_1 = \pi/2 E_J$; 
ii) Set $n_{x,a}^{-1}=n_{x,b}^{-1}= 1+ E_c^{(ab)} /2 E_{ch}$ and let
the system evolve for a time $t_1 = \pi / 4 E_{c}^{(ab)}$; 
iii) rotate again the $b$-island of an angle $3 \pi/4$
keeping it at degeneracy for a time $t_3 = 3\pi/2 E_J$. 
In the computational basis this corresponds to the operation
\begin{equation}
{\cal G}= \pmatrix{1&0&0&0 \cr 0&1&0&0 \cr 0&0&0&-i \cr 0&0&i&0}
\end{equation}
which is a C--Not gate apart from a phase factor. It is easy to verify that
the deviations from ideal gate operation due to the presence of the
electrostatic coupling during single qubit operations are $O (
E_{c}^{(ab)}/E_J)^2$. The gate ${\cal G}$ can be used to transform
entanglement into coherence of the control qubit, thus providing a
mechanism to discriminate entanglement from classical correlations.
Indeed we have:

\begin{eqnarray}
 {\cal G} |\Psi^{(-)}\rangle \langle \Psi^{(-)} | 
{\cal G} &\longrightarrow & \frac{1}{2} (|0\rangle - i|2\rangle)
(\langle 0|+ i \langle 2| )
\otimes |2\rangle \langle 2| \\
{\cal G} \rho_{mix} {\cal G} &\longrightarrow & \frac{1}{2} I
\otimes |2\rangle \langle 2|
\end{eqnarray}
Note that the target qubit ($b$) always factors, while $a$ is left
in either a coherent or a completely incoherent superposition of the basis
states, depending on the initial presence of entanglement. At this
point, one can reveal the coherence of the state of the control
qubit  by performing a current measurement as in~\cite{nakamura}.
If the excess current shows oscillations with respect to a varying
rotation angle, then the state is coherent and therefore the
initial state was a singlet. On the other hand,
if the excess current coming from island $a$ is constant one can
infer that the initial state was not entangled.

We mention that a Bell state analyzer can be easily implemented with
this setup. Indeed  by applying ${\cal G}$ followed by a further
$\pi/4$ rotation, the four maximally entangled Bell states,
$|\Phi^{(\pm)}\rangle = \frac{1}{\sqrt{2}} (|00\rangle \pm |22\rangle)$
and $|\Psi^{(\pm)}\rangle = \frac{1}{\sqrt{2}} (|02\rangle \pm
|20\rangle)$, are transformed into the basis state $|20\rangle,
|00\rangle, |02\rangle$ and $|22\rangle$, respectively, so that they
can be distinguished by two local single-qubit current (or charge) measurements.

All the preceding discussion could have been equally well phrased in terms 
of charge measurements instead of current ones. In this case, the right junctions 
are no longer necessary, while the remaining single Cooper pair
boxes should be coupled to two electrometers~\cite{delsing} which, in turn, 
induce a dephasing and a mixing among the qubit states~\cite{yuriy}. 
The procedure can be easily  generalized to include more qubits. 
With a three Josephson qubits setup, GHZ states of the form 
$\frac{1}{\sqrt{2}} (|000\rangle + e^{i
\chi} |222\rangle)$, \cite{ghz} can be obtained by applying two ${\cal G}$ 
gates, the first one to $b$ and $a$ and
the second one to $b$ and $c$ (with $b$ always as the control qubit). 

\acknowledgments The authors would like to thank M.-S. Choi, G. Falci, G. L. Ingold, Y. Makhlin, G. Sch\"on, J. Siewert, and A. Shnirman for helpful 
discussions. This works has been 
supported by the EU under IST-FET contracts EQUIP and SQUBIT,  by INFM
under contract PRA-SSQI, by Elsag S.p.A. and by ESF QIT Programme.

\begin{references}
\bibitem{nota}
    	In order to implement this second detection strategy,
    	we need $E_c^{ab}\gg \Gamma$,
    	but still $E_c^{ab}\ll E_J$.
\end{references}
\end{document}